\documentclass[iicol,sn-mathphys-num]{sn-jnl}

\usepackage{graphicx}%
\usepackage{multirow}%
\usepackage{amsmath,amssymb,amsfonts}%
\usepackage{amsthm}%
\usepackage{mathrsfs}%
\usepackage[title]{appendix}%
\usepackage{xcolor}%
\usepackage{textcomp}%
\usepackage{manyfoot}%
\usepackage{booktabs}%
\usepackage{algorithm}%
\usepackage{algorithmicx}%
\usepackage{algpseudocode}%
\usepackage{listings}%

\begin{document}

\title[Article Title]{Multimaterial Inkjet Printing of Mechanochromic Materials}

\author[1]{\fnm{Muriel} \sur{Mauron}}

\author[1]{\fnm{Lucie} \sur{Castens Vitanov}}

\author[1]{\fnm{César} \sur{Michaud}}

\author[1]{\fnm{Raphaël} \sur{Wenger}}

\author[1]{\fnm{Nicolas} \sur{Muller}}

\author[1]{\fnm{Roseline} \sur{Nussbaumer}}

\author[2]{\fnm{Céline} \sur{Calvino}}

\author[3,4]{\fnm{Christoph} \sur{Weder}}

\author[3,4,5]{\fnm{Stephen} \sur{Schrettl}}

\author*[1]{\fnm{Gilbert} \sur{Gugler}}\email{gilbert.gugler@hefr.ch}

\author*[3,4,6]{\fnm{Derek} \sur{Kiebala}}\email{derek.kiebala@uni-mainz.de}

\affil[1]{\orgdiv{iPrint Institute, HEIA-FR}, \orgname{HES-SO University of Applied Sciences and Arts Western Switzerland}, \orgaddress{\city{Fribourg}, \postcode{1700}, \country{Switzerland}}}

\affil[2]{\orgdiv{Freiburg Center for Interactive Materials and Bioinspired Technologies}, \orgname{University of Freiburg}, \orgaddress{\city{Freiburg}, \postcode{79110}, \country{Germany}}}

\affil[3]{\orgdiv{Adolphe Merkle Institute}, \orgname{University of Fribourg}, \orgaddress{\city{Fribourg}, \postcode{1700}, \country{Switzerland}}}

\affil[4]{\orgdiv{National Competence Center in Research Bio-inspired Materials}, \orgname{University of Fribourg}, \orgaddress{\city{Fribourg}, \postcode{1700}, \country{Switzerland}}}

\affil[5]{\orgdiv{TUM School of Life Sciences}, \orgname{Technical University of Munich}, \orgaddress{\city{Freising}, \postcode{85354}, \country{Germany}}}

\affil*[6]{\orgdiv{Department of Chemistry}, \orgname{Johannes Gutenberg University of Mainz}, \orgaddress{\city{Mainz}, \postcode{55128}, \country{Germany}}}

\abstract{Inkjet printing technology achieves the precise deposition of liquid-phase materials \textit{via} the digitally controlled formation of picoliter-sized droplets. Beyond graphical printing, inkjet printing has been employed for the deposition of separated drops on surfaces or the formation of continuous layers, which allows to construct materials gradients or periodic features that provide enhanced functionalities. Here, we explore the use of multinozzle, drop-on-demand piezoelectric inkjet technology for the manufacturing of mechanochromic materials, \textit{i.e.,} materials that change their color or fluorescence in response to mechanical deformation. To accomplish this, suitable polyurethane polymers of differing hardness grades were tested with a range of organic solvents to formulate low-viscosity, inkjet-printable solutions. Following their rheological characterization, two solutions comprised of “soft” and “hard” polyurethanes were selected for in-depth study. The solutions were imbibed with a mechanochromic additive to yield fluorescent inks, which were either dropcast onto polymeric substrates or printed to form checkerboard patterns of alternating hardness using a lab-built, multimaterial inkjet platform. Fluorescence imaging and spectroscopy were used to identify different hardness grades in the dropcast and printed materials, as well as to monitor the responses of these gradient materials to mechanical deformation. The insights gained in this study are expected to facilitate the development of inkjet-printable, mechanochromic polymer materials for a wide range of applications.}


\maketitle
\clearpage
\section{Introduction}\label{intro}

Voxelated matter, \textit{i.e.,} materials comprised of modular, three-dimensional building blocks, has attracted growing interest due to the fact that such materials can be designed to exhibit tailored surface functionalities and gradient properties \cite{Lewis.2019, doubrovski_voxel-based_2015}. Inkjet-based 3D printing is the method of choice for fabricating voxelated materials with high precision \textit{via} the digitally controlled formation of picoliter-sized droplets, affording access to a range of materials functionalities otherwise inaccessible with conventional processing methods \cite{Lewis.2019}. For example, it has been shown that cell growth and stem cell differentiation can be influenced by printing gradients of biologically active materials \cite{ilkhanizadeh_inkjet_2007, cai_inkjet_2009}, and biomedical tomographic data sets have been converted into topographical models \textit{via} inkjet gradient printing \cite{hosny_improved_2018}. Beyond biomedical applications, gradients of functional or mechanical characteristics have been demonstrated in 3D-printed metals \cite{godlinski_steel_2005}, ceramics \cite{mott_zirconiaalumina_1999, afzal_functionally_2012,lee_digital_2021}, as well as polymers \cite{Schaffner2018}. This has been achieved by, for example, reactive inkjet printing in which multiple printheads eject pre-polymer reactants onto the same substrate, which form the desired material \textit{in situ} with well-resolved microscale features \cite{schuster_reactive_2019}. Moreover, advancements in printhead architecture, optionally assisted by microfluidic mixers, have enabled the printing of materials with gradient properties \textit{via} a technique known as multimaterial, multinozzle 3D printing, which has greatly expanded the potential for generating new types of inkjet-printed materials \cite{Lewis.2019, doubrovski_voxel-based_2015,hardin_microfluidic_2015}.

\begin{figure*}[h]
\centering
\includegraphics[width=0.9\textwidth]{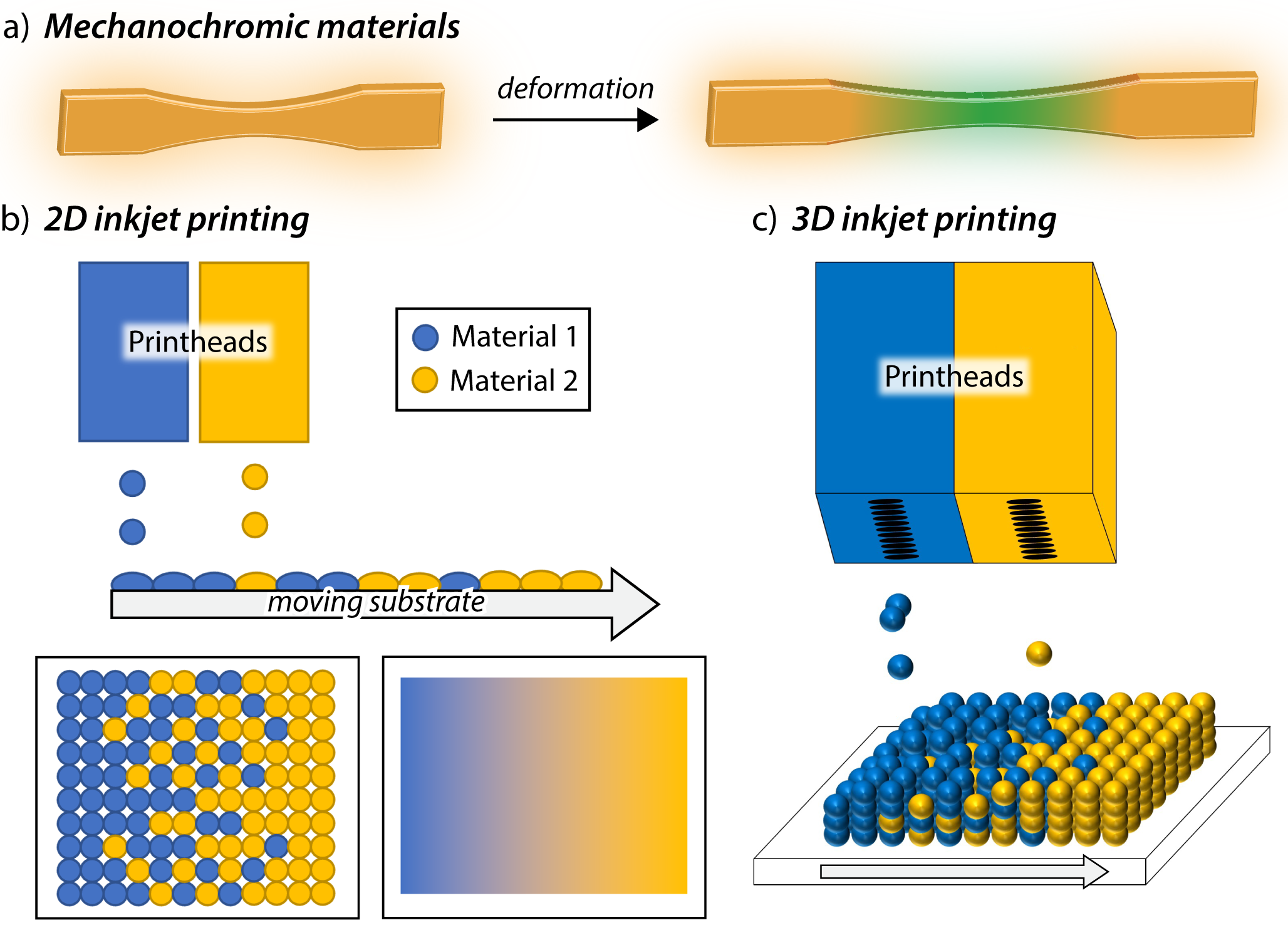}
\caption{a) Operating principle of mechanochromic materials. When subjected to mechanical deformation, the color and/or fluorescence of the material changes from, for example, orange to green. b-c) Overview of how inkjet printing is used to digitally control the composition of a continuous gradient between two materials. b) \textit{Top:} Side view of a line printed with two printheads each jetting a different polymer-based ink. The ink deposited at each location is digitally controlled by selecting which printhead jets a drop as the substrate passes underneath them. \textit{Bottom left:} Simplified top-view of a 2D surface printed with two printheads to achieve a gradient. \textit{Bottom right:} Industrial printheads typically feature hundreds to thousands of nozzles, allowing for much smoother gradients. c) Schematic depicting the deposition of multiple layers on top of each other to create a three-dimensional gradient material.}\label{fig1}
\end{figure*}

Here, we sought to harness the potential of multimaterial printing to fabricate gradient property materials endowed with mechanochromic functionality, \textit{i.e.,} the ability to change their color or fluorescence in response to mechanical deformation (Fig.~\ref{fig1}a). Such mechanochromic materials are particularly useful for reporting stresses and damage in load-bearing materials \textit{via} an easily detectable and in some cases visually discernible optical signal \cite{Traeger2020}. The fabrication of such mechanoresponsive polymer materials is made possible by the incorporation of a mechanophore, namely a molecular entity that produces a defined response to external forces \textit{via} the breakage of a labile bond \cite{Deneke2020}. Recently, some of us developed a macromolecular additive (tOPV) consisting of two excimer-forming oligo(\textit{p}-phenylene-vinylene) (OPV) dyes connected \textit{via} a telechelic poly(ethylene-\textit{co}-butylene) backbone ($M_n$ = 3,000 g$\cdot$mol$\cdot$L$^{-1}$) that is particularly convenient for conferring mechanochromic behavior to a wide range of polymers \cite{Calvino2019}. Already when blended in very small quantities (\textit{i.e.,} 0.2–1.0 wt\%) into a host polymer, tOPV phase-separates from the matrix to form microscopic, fluorescent inclusions whose mechanical and optical properties are closely tied to that of the host matrix, while leaving the mechanical properties of the latter unaffected \cite{kiebala_submicrometer_2023}. This blending process has been used to fabricate films comprised of polyurethane (PU), polyisoprene (PI), poly(styrene-\textit{b}-butadiene-\textit{b}-styrene) (SBS) and poly($\epsilon$-caprolactone) (PCL) that display strain-dependent changes in fluorescence color \cite{Calvino2019,kiebala_submicrometer_2023}. In fact, a change in emission wavelength of a tOPV-containing elastomeric polyurethane was spectroscopically discernible at strains as low as 5\% \cite{Kiebala2020}, which opens the possibility for sensing applications with unprecedented precision. Taken together, the advantages offered by this mechanochromic additive make it an ideal candidate for exploring the fabrication of mechanochromic materials \textit{via} inkjet printing.

Thus, to pursue the inkjet printing of gradient-property, mechanochromic materials, we employed a multimaterial printing system in which two printheads are mounted in parallel and filled with inks containing polymers with two different hardness grades. Droplets of both inks were printed in an alternating fashion on a substrate, enabling the creation of a variable hardness, multilayer material (Fig. \ref{fig1}b-c). Imbibing the polymer inks with the aforementioned mechanochromic additive was possible without the need for additional pre-processing or chemical modification, in contrast to polymer inks formulated with mechanochromic nanoparticles \cite{ogumi_inkjet_2022} or microgels \cite{xu_multi-material_2023}. This yielded patterned fluorescent films, which allowed us to investigate the mixing dynamics of the different materials at the voxel-voxel interface using fluorescence imaging and spectroscopy. Finally, we show that mechanical deformation of the substrate-supported, gradient-hardness films changes their fluorescence in a quantifiable fashion, constituting an important step towards the realization of inkjet-printed, mechanochromic multimaterials.

\section{Materials and methods}\label{methods}

\subsection{Ink formulation}\label{methods-sub1}

As the basis for the mechanochromic polymer inks formulated in this study, commercially available polyurethanes (PUs) of different hardness, namely ``soft" PU (PU-S, shore hardness 35A) and ``hard" PU (PU-H, shore hardness 80A) were provided by BASF and Huntsmann, respectively. 1,4-Dioxane solvent (also referred to as dioxane; 99\%+ purity) was purchased from Acros Organics. The solvents 1-methyl-2-pyrrolidinone (NMP; 99\%) and 2-butoxyethyl acetate (BCA; 99\%) were purchased from Sigma Aldrich. The telechelic oligo(\textit{p}-phenylene-vinylene) (tOPV) mechanochromic additive used in the final ink formulations was prepared as previously reported, omitting the final purification step \cite{Calvino2019}. Hence, a greater amount of tOPV was used for the ink formulations (\textit{i.e.,} 5 wt\% vs. polymer) in comparison to previous studies (\textit{i.e.,} 0.2–1 wt\% vs. polymer) to ensure sufficient assembly of difunctional tOPV in the printed samples.

Chemical compatibility tests were performed between the candidate solvents and the printheads following the manufacturer’s instructions. To carry out these tests, nine different parts of the printhead were weighed, immersed individually in 30 mL of the solvent being tested in amber glass bottles, and placed in an oven at 60~$^{\circ}$C for six weeks. At two-week intervals, the parts were cleaned by immersion first in \textit{n}-butyl lactate (99\%, ThermoScientific) and then in propan-2-ol (99\%, Thommen-Furler AG). After evaporation of the cleaning solvents, the parts were re-weighed and were again immersed in the test solvent at 60~$^{\circ}$C. The data was sent back to the manufacturer, who analyzed the changes in weight of each piece to assess compatibility between the solvent and the printhead.

The rheological properties of the polymer solutions were analyzed with a Piezo Axial Vibrator (Tri-PAV) rheometer from TriJet at room temperature from 100 Hz to 10 kHz to identify suitable solvents and polymer concentrations for inkjet printing \cite{hussain_high-frequency_2022}. The surface energy, contact angle and surface tension of the solutions were measured with the OCA25 contact-angle and drop-shape measuring device from Dataphysics. The surface energy of the substrate was calculated using the contact angles of droplets ($V \approx 2~\mu$L) of water and ethylene glycol, the Young Laplace equation, and the Owens-Wendt-Rabel and Kaelble (OWRK) model. The static surface tension for droplets ($V = 12$–$15~\mu$L) of each ink was measured with the same device using the pendant drop method.

\subsection{Inkjet printing}\label{methods-sub2}

To enable simultaneous printing of multiple different PU inks, an inkjet printing platform was built that incorporates two Seiko RC1536L printheads, an ink distribution system, and a stage movable along \textit{x}-, \textit{y}-, and \textit{z}-axes, all of which is governed by a Beckhoff Programmable Logic Controller (PLC) (Fig. A1). The Seiko RC1536L printheads, which features four rows of 384 nozzles (totaling 1536 available nozzles with a maximum resolution of 360 dpi), were selected for their generally good chemical compatibility and suitability for printing at relatively high viscosity (\textit{i.e.,} up to 20~mPa$\cdot$s). The platform is capable of simultaneously printing materials in 2D with up to three printheads, as well as 3D materials by sequentially printing multiple 2D layers as depicted in Fig. \ref{fig1}c. The volume and speed of ink droplets exiting the printhead nozzles during the printing process were determined by acquiring images of the droplets in flight using a camera (33GP031 from The Imaging Source) with a 0.7–4.5X zoom lens (Model HY-180XA from Hayear), combined with a strobing light source and a homemade triggering system. A bespoke Matlab script was used for synchronizing the operations and extracting key features from the images, such as the drop speed, volume, and shape. The printheads and their waveforms were controlled with driving electronics from Aewa and their dedicated APRINT software. Selected inks containing the mechanochromic additive were deposited by dropcasting or inkjet printing at room temperature on a PU substrate fixed on a heating plate at 40~$^{\circ}$C, followed by drying at 60~$^{\circ}$C for 12 h. Round substrates (55 mm diameter, 0.3 mm thickness) were produced by Torson Injex by injection molding of Elastollan E565A12P (BASF), and rectangular substrates (38 × 20 × 60 mm) of the same material were produced by injection molding at the iRap Institute of the HEIA in Fribourg.

\subsection{Fluorescence spectroscopy and microscopy}\label{methods-sub3}

Fluorescence changes in printed PU/tOPV films were measured spectroscopically as a function of mechanical stress using an Ocean Optics USB 4000 spectrometer connected to an Ocean Optics LS-450 LED light source with an excitation wavelength of $\lambda\mathrm{_{ex}}$ = 380 nm and an Ocean Optics QR230-7-XSR SMA 905 optical fiber. The extent of aggregation of the excimer-forming end groups of the tOPV macromolecule, which is influenced by processing conditions, changes in temperature, and the application of mechanical force, were probed by measuring fluorescence spectra and determining the monomer-to-excimer emission intensity ratio (\textit{I}$_{510}$/\textit{I}$_{630}$). Mechanical deformation of tOPV-containing materials has been shown to reliably result in an increase in \textit{I}$_{510}$/\textit{I}$_{630}$, which occurs when the distance between adjacent tOPV emitters is increased \cite{Kiebala2020,kiebala_submicrometer_2023}. The value was calculated from the solid-state fluorescence spectra by taking the intensity value at the maximum of the monomer emission peak ($\lambda\mathrm{_{max}}$ = 510 nm) divided by the maximum of the excimer emission peak ($\lambda\mathrm{_{max}}$ = 630 nm). Samples were placed on top of a piece of black paper, and the optical fiber was oriented normal to the surface at a distance of 2 mm, resulting in a spot size of approximately 800 $\mu$m from which the diffuse reflectance was measured. Spectra were recorded before, after, or during application of mechanical stress and fluorescence data were acquired using Stream Basic software. Stress-strain data collected during these measurements was recorded by uniaxial deformation of rectangular samples with dimensions of 30 × 5 × 1 mm (length × width × thickness) at 100\% min$^{-1}$ using a Linkam TST350 microtenstile stage equipped with a 20 N load cell and controlled by the accompanying Linksys32 software. A pre-load force of 0.1–0.3 N was applied before initiating deformation.

Confocal microscopy images were acquired on a Zeiss LSM 710 Meta confocal laser scanning microscope (CLSM) equipped with an Ar laser (max. power 25 mW) and a spectrometer allowing acquisition of emitted light in two channels at nanometer precision within the visible range. All data was recorded using a 63X/1.3NA oil lens at a lateral resolution of 132 nm. Channel 1 recorded emission between 461 and 525 nm (\textit{i.e.,} primarily tOPV monomer emission), and channel 2 recorded emission between 550 and 725 nm (\textit{i.e.,} primarily tOPV excimer emission). Both channels used a laser excitation of 458 nm. For a given region of interest (ROI) at the surface of the sample, the monomer-to-excimer ratio (for CLSM images denoted as \textit{I}$\mathrm{_{M}}$/\textit{I}$\mathrm{_{E}}$) was obtained by dividing the mean pixel intensity for the ROI in channel 1 by the mean intensity for the ROI in channel 2. While \textit{I}$\mathrm{_{M}}$/\textit{I}$\mathrm{_{E}}$ values obtained by CLSM image analysis are not directly comparable to the \textit{I}$_{510}$/\textit{I}$_{630}$ values obtained by solid-state fluorescence spectroscopy, the trends in these values (\textit{e.g.,} for PU of different hardness, or when subjecting the films to mechanical force) mirror each other. CLSM images were acquired as 16-bit grayscale images, and false colors of green and orange were applied to the monomer and excimer channels, respectively, to match the visible fluorescence colors of the corresponding assembly states of tOPV. Thus, combined-channel CLSM images have a yellow-green appearance after false coloring.

Printed PU/tOPV layers on Elastollan substrates were imaged by cutting a 0.5 × 0.5 cm square piece and placing it on a glass slide. Two drops of halogen-free, non-fluorescent Cargille Immersion Oil Type HF (\textit{i.e.,} refractive index-matching oil) were added to the sample, onto which a glass cover slip of approximately 0.17~mm thickness was placed. One drop of the same immersion oil was then placed on the lens of the confocal microscope, after which the glass slide and sample assembly was inverted such that the cover slip was facing downwards and placed on the microscope stage. The lens was then raised so that the oil came into contact with the glass cover slip, creating an air-free path for the excitation laser of the laser confocal scanning microscope to pass through the sample. Equibiaxial stretching was carried out on samples using a custom-built stretching device that applied radial strain in all directions (see Ref. \cite{kiebala_submicrometer_2023} for a detailed description of the setup).

\section{Results and discussion}\label{results}

\subsection{Polymer solution testing}\label{results-sub1}

As a first step in the formulation of printable mechanochromic inks, solubility tests were carried out with the PU polymers and the tOPV mechanochromic additive in a range of organic solvents with varying hydrophobicity and boiling points. Of the solvents tested, tOPV showed the best solubility at 1 mg/mL in dioxane, tetrahydrofuran (THF), \textit{N}-methyl-2-pyrrolidinone (NMP), and mixtures of NMP and 2-butoxyethyl acetate (BCA) with at least 25 wt\% of NMP. tOPV remained well-dissolved in these solutions at room temperature over the course of 24 h as indicated by the solutions' green fluorescent color, thus serving as a good indicator that no tOPV would precipitate inside the printhead during the printing process. The other solvents tested that did not fully dissolve the PU (in the range of 15–20 mg/mL) or tOPV (at 1 mg/mL) include: \textit{n}-hexane, xylene, isophorone, BCA (99\%), methyl ethyl ketone, ethyl acetate, methanol, ethanol, propanol, 1,5-pentanediol, glycerol, ethylene glycol, propylene glycol, 2-methoxyethanol, and triethanolamine.

\begin{figure}[h]
\centering
\includegraphics[width=0.45\textwidth]{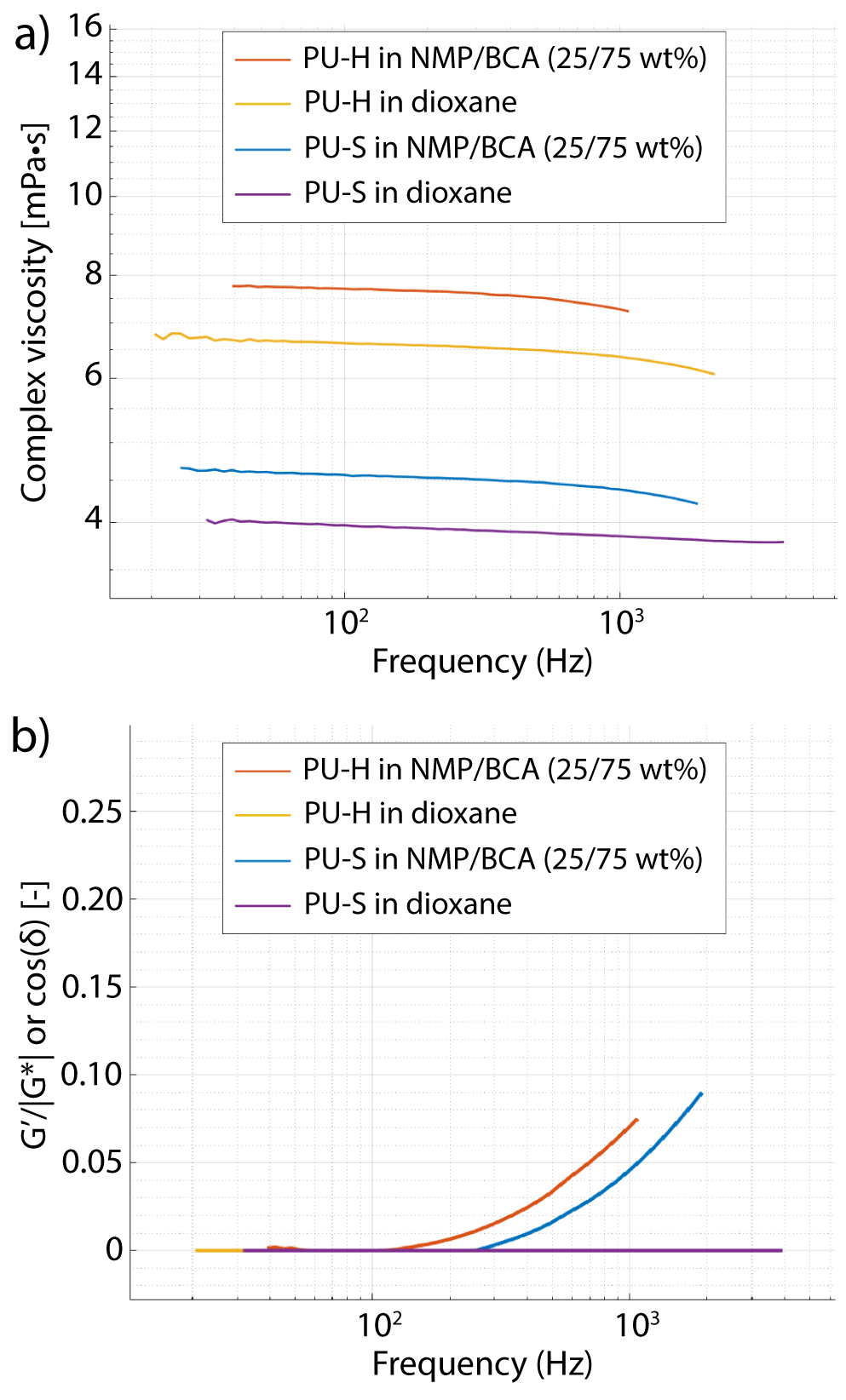}
\caption{Rheological measurements carried out on PU-S and PU-H at 20 mg/mL in both dioxane and NMP/BCA 25/75 wt\% solvents. Both (a) complex viscosity and (b) elasticity (G’/$|$G*$|$) were measured as a function of applied frequency at 25~$^{\circ}$C.}\label{fig2}
\end{figure}

\begin{figure}[h]
\centering
\includegraphics[width=0.45\textwidth]{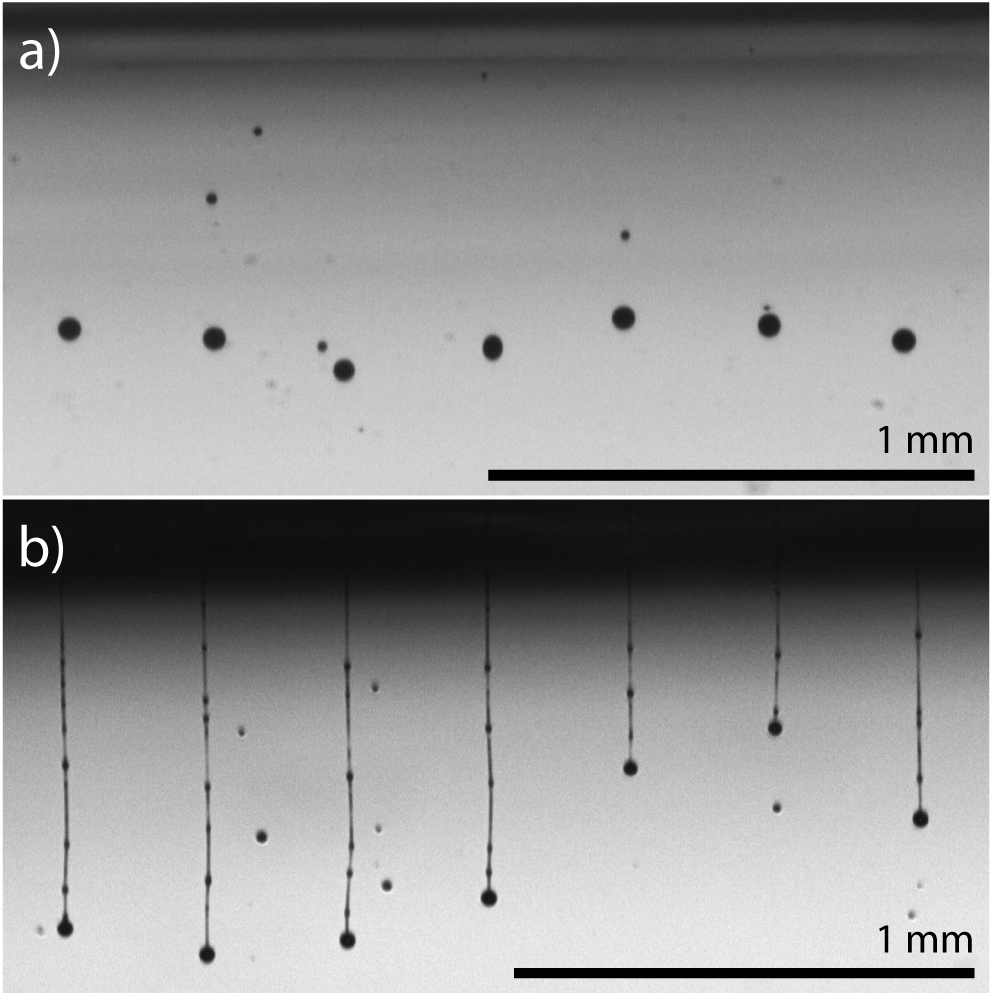}
\caption{Real-time images of polymer solution droplets ejected from seven adjacent nozzles in one row of the printhead at (a) 20~V ($v_{drop}\approx$ 5~m/s) and at (b) 28~V ($v_{drop}\approx$ 8~m/s). Scale bars = 1 mm.}\label{fig3}
\end{figure}

After subjecting the different components of the Seiko RC1536L printhead used in the lab-built inkjet printer to chemical compatibility tests (see Section \ref{methods-sub1} for details), the results showed dioxane and NMP/BCA 25/75 wt\% to be the most promising solvent candidates (Table A1). Polymer solutions were then made by dissolving pellets of PU-S and PU-H in dioxane and NMP/BCA 25/75 wt\% at 20 mg/mL, which was the concentration used for all further tests. To assess the printability of the different ink formulations, rheological measurements were carried out on solutions of PU-S and PU-H without the tOPV additive, given that the small amounts of tOPV to be added in the final inks would not affect the rheological properties of the latter. The results showed that all four polymer solutions exhibit complex viscosities ranging from 3.5–8 mPa$\cdot$s between 20–3000 Hz applied frequency (Fig.~\ref{fig2}a), well within the range of 2–20 mPa$\cdot$s recommended for industrial printheads \cite{zapka_handbook_2017,zapka_measurement_2022}. Moreover, the elasticity of the solutions, defined as the ratio of the storage modulus G’ to the absolute value of the complex modulus $|$G*$|$, was found to be sufficiently low, \textit{i.e.,} near zero for the dioxane-based solutions and less than 0.1 for NMP/BCA 25/75 wt\% up to 3~kHz (Fig.~\ref{fig2}b), indicating that they could be reliably jetted through the printhead nozzle. Thus, dioxane was selected as the preferred solvent to formulate the mechanochromic inks based on the lower elasticity of the dioxane-based PU solutions. Finally, printhead waveform optimization was performed by acquiring and analyzing real-time images of polymer solution droplets ejected from the printhead nozzles (Fig.~\ref{fig3}) and iterating over the waveform parameters until suitable drop characteristics were achieved. The waveforms of the Seiko RC1536L printheads were initially optimized to achieve a minimum drop speed of 5 m/s and a drop size of 25 pL with the selected solutions. However, the quick drying of the solution in the nozzles prevented stable printing. Thus, a higher waveform voltage was employed (\textit{i.e.,} \textit{ca.} 28V instead of the 20 V initially used) to produce faster-traveling droplets ($v_{drop}\approx$ 8~m/s) that avoided precipitation of the polymer in the printhead during printing.

\subsection{Film-substrate interactions}\label{results-sub2}

Next, we investigated the interaction of the dioxane polymer solutions with an Elastollan polymer film (thickness = 1 mm), which was to be used as the substrate for dropcasting and ultimately printing the inks. To this end, we deposited a 10 $\mu$L drop of each solution onto the substrate and measured the surface tension of the solution droplet, the solution-substrate contact angle, and the surface energy of the substrate. The dioxane-based polymer solutions were found to exhibit a low static surface tension typical of solvent-based inks, namely 33.2 and 25.6 mN/m for PU-S and PU-H in dioxane, respectively, both of which fall below the 40 mN/m limit for inkjet drop ejection \cite{he_roles_2017}. The surface energy of the substrate was found to be 33.9 mN/m, which exceeds the surface tension and indicates excellent wettability. Finally, the contact angles were determined to be 21.8 $^{\circ}~\pm$ 0.9 $^{\circ}$ and 23.3 $^{\circ}~\pm$ 2 $^{\circ}$ for PU-S and PU-H in dioxane, respectively, further confirming that these inks wet the substrate well.

\subsection{Mechanochromism of dropcast films}\label{results-sub3}
\begin{figure}[h]
\centering
\includegraphics[width=0.45\textwidth]{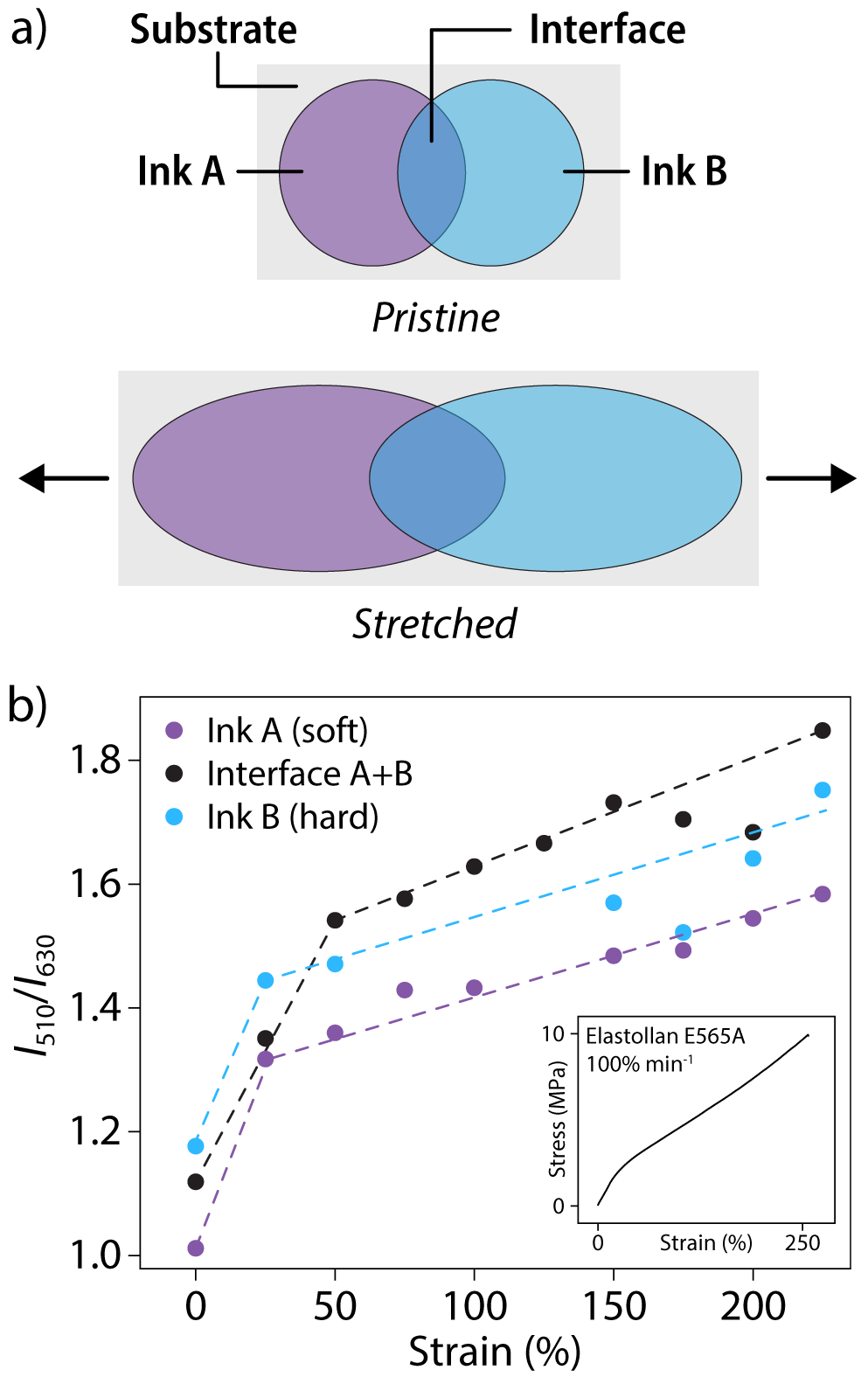}
\caption{a) Schematic depicting films made by dropcasting inks A and B onto the Elastollan substrate. \textit{Upper panel:} Regions of pure ink A and ink B are indicated, as well as the mixed region or interface where the two inks overlap. \textit{Lower panel:} Uniaxial stretching of the substrate caused the dropcast films to deform along the stretching axis. b) Plot of the monomer-to-excimer fluorescence ratio (\textit{I}$_{510}$/\textit{I}$_{630}$) of the pure and mixed ink regions as a function of applied strain. Dashed lines serve as a guide to the eye. \textit{Inset:} Stress-strain curve of the Elastollan substrate measured during the test.}\label{fig4}
\end{figure}

To endow these printable polymer solutions with mechanochromic properties, the tOPV additive was dissolved in each solution at a concentration of 1 mg/mL to produce the final ink formulations (\textbf{Ink A}: PU-S [20 mg/mL] + tOPV [1 mg/mL] in dioxane, and \textbf{Ink B}: PU-H [20 mg/mL] + tOPV [1 mg/mL] in dioxane). The inks were dropcast onto Elastollan substrates and subsequently dried at 60~$^{\circ}$C for 12 h \textit{in vacuo} to yield thin films that exhibited a homogeneous yellow fluorescent color under UV illumination, indicating proper assembly of the tOPV additive within the films \cite{Calvino2019,Kiebala2020}. The two inks were dropcast next to each other on the substrate in such a way that produced a “mixed region” or interface where the two inks overlapped (Fig.~\ref{fig4}a).

Solid-state fluorescence spectroscopy measurements revealed that, in the as-prepared films, the monomer-to-excimer ratio of ``soft" ink A (\textit{I}$_{510}$/\textit{I}$_{630}$ = 1.02) was lower than that of ``hard" ink B (\textit{I}$_{510}$/\textit{I}$_{630}$ = 1.18), and that of the mixed region (\textit{I}$_{510}$/\textit{I}$_{630}$ = 1.13) was in between the two. Next, the substrate was subjected to uniaxial tensile deformation at a strain rate of 100\% min$^{-1}$ up to 250\% strain (the resulting stress-strain curve is shown in the inset in Fig.~\ref{fig4}b), and the \textit{I}$_{510}$/\textit{I}$_{630}$ for each region of the dropcast film was monitored. Notably, the \textit{I}$_{510}$/\textit{I}$_{630}$ for all regions (\textit{i.e.,} pure inks A and B, as well as the mixed region) increased sharply between 0-50\% strain, after which the \textit{I}$_{510}$/\textit{I}$_{630}$ of each region continued to increase in a linear fashion until the end of the test (Fig.~\ref{fig4}b). This linear fluorescence response to uniaxial deformation closely mirrors that of previously reported PU/tOPV blend films that feature different slopes for low- vs. high-strain deformation \cite{Kiebala2020}, thus confirming that force is transferred between the substrate and dropcast films. Notably, a greater initial increase in \textit{I}$_{510}$/\textit{I}$_{630}$, as well as a slightly steeper slope at high strains, is observed for the interfacial region, suggesting that microscale domain boundaries between partially demixed PU-S and PU-H exert more local strain on the tOPV additive when deformed. Taken together, these results confirm that the PU/tOPV inks formulated herein can be readily used to fabricate substrate-supported, mechanochromic films.

\begin{figure*}[h]
\centering
\includegraphics[width=0.9\textwidth]{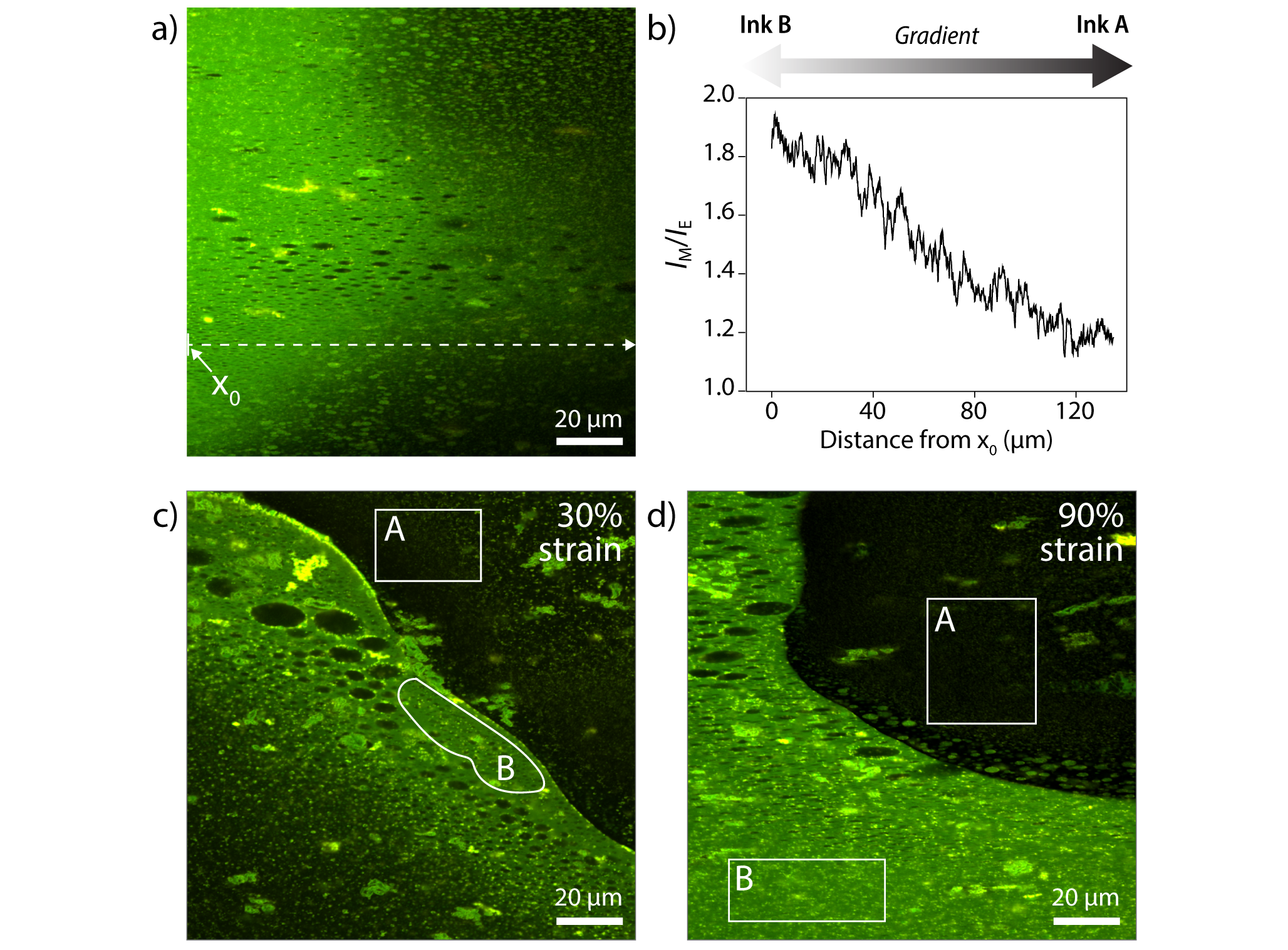}
\caption{a) Combined-channel 2D CLSM image of an interfacial region between inks A and B on the same film. b) \textit{I}$\mathrm{_{M}}$/\textit{I}$\mathrm{_{E}}$ values determined as a function of \textit{x}-position moving left to right along the dashed line in panel (a). c,d) Confocal microscopy images of the mixing interface of inks A and B, which were dropcast onto an Elastollan substrate and subjected to equibiaxial strains of (c) 30\% and (d) 90\%. Images were recorded in two channels, \textit{i.e.,} the monomer (Ch 1: $\lambda\mathrm{_{em}}$ =  461–525 nm) and excimer (Ch2: $\lambda\mathrm{_{em}}$ = 550–725 nm) emission channels and then merged to obtain the combined-channel images shown. At each strain, the monomer-to-excimer emission ratio (\textit{I}$\mathrm{_{M}}$/\textit{I}$\mathrm{_{E}}$) was determined for the indicated regions of interest for inks A and B (labeled as A and B, respectively) by dividing the mean pixel intensity of channel 1 by that of channel 2. Scale bars = 20 $\mu$m.}\label{fig5}
\end{figure*}

After confirming the mechanochromism of dropcast films made from polymer inks A and B, we turned to confocal laser scanning microscopy (CLSM) to more closely study the morphology of the films, particularly at the mixing interface. To this end, CLSM images were acquired. The monomer and excimer emission of the dropcast films were recorded in two separate channels, thus allowing for the determination of the monomer-to-excimer emission ratio (denoted as \textit{I}$\mathrm{_{M}}$/\textit{I}$\mathrm{_{E}}$ for CLSM images) with micrometer-resolution for different areas of the film \textit{via} image analysis (see Section \ref{methods-sub3} for details). Distinct areas comprised of inks A and B were readily identifiable by their fluorescence; namely, the area formed by “soft” ink A consistently exhibited a lower \textit{I}$\mathrm{_{M}}$/\textit{I}$\mathrm{_{E}}$ than regions formed by “hard” ink B (Fig.~\ref{fig5}). In some regions, a quasi-linear decrease in \textit{I}$\mathrm{_{M}}$/\textit{I}$\mathrm{_{E}}$ is observed across the mixing zone when moving from ink B (left side of the image in Fig.~\ref{fig5}a) to ink A (right side), indicative of a compositional gradient between the two inks (Fig.~\ref{fig5}a,b). On the other hand, some areas of the mixing zone exhibit a sharp transition between the two inks (Fig.~\ref{fig5}c,d, Fig. A2), likely stemming from limited miscibility of the two polymers and selective dewetting processes that occurred during drying. Moreover, low-\textit{I}$\mathrm{_{M}}$/\textit{I}$\mathrm{_{E}}$ features were observed interspersed in regions of ink B and vice versa, pointing to a complex mixing behavior between the two inks (Fig.~\ref{fig5}a,c,d).

Next, the dropcast films were subjected to equibiaxial strain (\textit{i.e.,} radial strain applied equally in all directions) using a custom-built stretching device (see Ref. \cite{kiebala_submicrometer_2023} for details) and imaged \textit{via} CLSM. At each strain, parts of film comprised of either ink A or B were identified, and the \textit{I}$\mathrm{_{M}}$/\textit{I}$\mathrm{_{E}}$ was determined for the indicated regions by dividing the mean pixel intensity of the monomer channel by that of the excimer channel. The results show that, when increasing the equibiaxial strain from 30\% to 90\%, the \textit{I}$\mathrm{_{M}}$/\textit{I}$\mathrm{_{E}}$ for ink A increases from 0.98 to 1.09, and the \textit{I}$\mathrm{_{M}}$/\textit{I}$\mathrm{_{E}}$ for ink B increases from 1.47 to 1.57 (Fig.~\ref{fig5}c,d). These observations are in agreement with the increase in \textit{I}$\mathrm{_{510}}$/\textit{I}$\mathrm{_{630}}$ measured for both inks A and B \textit{via} solid-state fluorescence spectroscopy (Fig.~\ref{fig4}b) and further corroborate that PU/tOPV films formed from the dioxane-based inks formulated herein exhibit mechanochromism.

\subsection{Inkjet printing trials}\label{results-sub4}
\begin{figure}[h]
\centering
\includegraphics[width=0.475\textwidth]{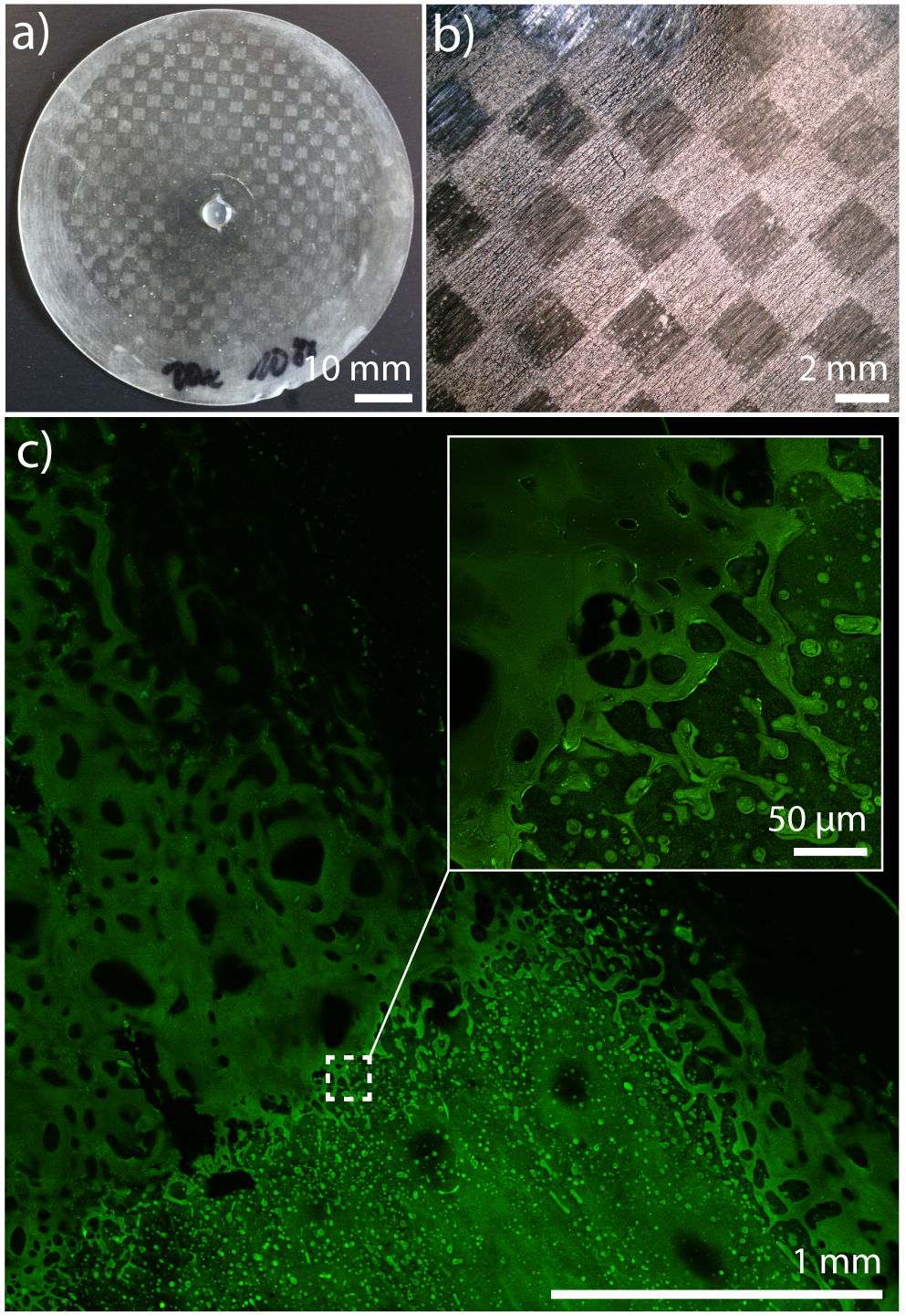}
\caption{a) Image of a checkerboard pattern created by printing polymer solutions of “soft” PU-S and “hard” PU-H (\textit{c}(dioxane) = 20 mg/mL) in an alternating fashion on a circular Elastollan substrate. Multiple passes were made by the printhead to deposit a total of 20 layers. Scale bar = 10 mm.  b) Magnified view of a portion of the sample shown in (a). Scale bar = 2 mm. c) Combined-channel CLSM image of the intersection of two of the checkerboard squares comprised of “soft” ink A (lower right region) and “hard” ink A (upper left region) that had been printed on the same type of circular Elastollan substrate as shown in (a). Scale bar = 1 mm. Inset: Magnified CLSM image of the indicated interfacial region between the two inks (scale bar = 50 $\mu$m).}\label{fig6}
\end{figure}

Finally, we sought to employ our inkjet printing setup to fabricate mechanochromic films with gradient mechanical properties. As a first test, polymer solutions A and B (\textit{i.e.,} inks A and B without the tOPV additive) were printed in an alternating fashion onto a circular Elastollan substrate to create a checkerboard pattern (Fig.~\ref{fig6}a,b). Multiple passes were made by the printhead to deposit a total of 20 layers, and a feature resolution of 1 mm was obtained in the \textit{x,y}-plane as seen by the naked eye. The same process was then carried out with tOPV-containing inks A and B, and the 20-layer printed pattern was imaged by CLSM (Fig.~\ref{fig6}c). While the interface between the “soft” ink A and “hard” ink B is clearly discernible, many satellite droplets of the hard PU are visible at least 0.5 mm into the soft PU region. This is largely a consequence of the fact that, in order to avoid the inks drying in the printhead during deposition, a higher waveform voltage was used to increase the speed of the droplet ejection. Thus, the CLSM images of the additive-containing, printed PU patterns reveal that the high voltage and speed of the printing process diminish the accuracy and therefore resolution of the printed features.  Moreover, many non-fluorescent areas are visible within the hard PU region, indicating that this PU shows a greater tendency to dewet from the substrate during printing and leave some portions of the substrate uncovered. To overcome these obstacles, further trials are needed to identify alternative polymer materials that are compatible with inkjet-printable solvents and that, like films printed from ink B, do not show evidence of dewetting from the substrate when multiple-pass printing is performed. Moreover, the use of a less volatile solvent could allow for slower printing, which may avoid the generation of satellite droplets that breach the hard-soft polymer interface.

\section{Conclusion}\label{conclusion}
In the present work, the printing of gradient-property, patterned polymer films with mechanochromic sensing capability was investigated. In order to realize the fabrication of such films, a multimaterial inkjet printing platform was developed that allows for the simultaneous printing of two different polymer solutions into multilayer voxels of alternating composition. After extensive testing was carried out on candidate polymers solutions to evaluate their printability, “hard” and “soft” PU solutions in dioxane were selected for their desirable rheological properties and compatibility with the printheads. A mechanochromic additive was mixed into the solutions to obtain fluorescent polymer inks, which were either dropcast or printed in different patterns onto a stretchable polymer substrate. The different inks were readily identifiable \textit{via} their distinct fluorescence, and an in-depth evaluation of the microscale features of the fluorescent films shed light on complex mixing behaviors at the interface between the two inks. Importantly, when subjected to mechanical deformation, both dropcast and inkjet-printed, multimaterial films exhibited mechanochromic behavior that could be tracked individually for each voxel. The insights gained into the properties of inkjet-printable polymer inks and the microscale characterization of patterned polymers are expected to greatly facilitate the development of tailor-made functional materials with complex property profiles otherwise inaccessible by conventional processing techniques.

\backmatter

\bmhead{Supplementary information}

Electronic Supplementary Information (ESI) available: Supplementary Figures A1-A2 and Supplementary Table A1.

\bmhead{Acknowledgements}

M.M., L.C.V., C.M., R.W., R.N., N.M. and G.G. acknowledge financial support through funding from HES-SO University of Applied Sciences and Arts Western Switzerland, Engineering and Architecture, Grant SmartMatJet 114624. S.S. and D.K. gratefully acknowledge financial support through the National Center of Competence in Research (NCCR) Bio-inspired Materials, a research instrument of the Swiss National Science Foundation (SNF), and funding from the Adolphe Merkle Foundation.

\section*{Statements and Declarations}

\bmhead{Author contributions}
M.M., L.C.V., C.M., and R.W. built the inkjet printing device; carried out the printhead compatibility testing, polymer solution characterization, and substrate analysis; optimized the dropcasting protocol; and carried out the inkjet printing trials. D.K. carried out solubility testing and prepared the mechanochromic inks, measured the fluorescence of dropcast films during tensile testing, and acquired and analyzed the confocal microscopy images of printed films. C.C. synthesized the mechanochromic additive. S.S. and G.G. designed the original concept for the study and provided guidance throughout the project. D.K., M.M., R.N., R.W., and N.M. wrote the manuscript, which was edited by C.C. All authors have given approval to the final version of the manuscript.

\bmhead{Data availability}
The datasets generated and analyzed during the current study are available from the corresponding author on reasonable request. The source data generated during this study will be deposited to the Zenodo repository after acceptance of the article.

\bmhead{Competing interests}
The authors declare no competing financial interests.



\appendix
\begin{appendices}
\clearpage
\section{}\label{sup-info}

\begin{table*}[h]
\caption{Results of the compatibility tests carried out with different parts of the Seiko RC1536L printhead and the three candidate solvents dioxane, NMP/BCA 25/75 wt\%, and NMP/BCA 50/50 wt\%. Changes in the mass of each printhead component due to solvent uptake or degradation the components are reported as percentages (positive for increased mass, negative for reduced mass) after immersing the components in the indicated solvent for 6 weeks at 60 $^{\circ}$C. Bold values fall outside of the manufacturer’s specification range and thus indicate insufficient compatibility.}\label{tableS1}%

\begin{center}
\begin{tabular}{ |p{1cm}|p{2cm}|p{2cm}|p{3cm}|p{3cm}|  }
\toprule
Part & Material & \multicolumn{3}{|c|}{Mass increase (wt \%) in:}\\
\midrule
\multicolumn{2}{|c|}{} & Dioxane & NMP/BCA~(25/75) & NMP/BCA~(50/50) \\
\midrule
1   & Plastic       & 1.1     & 0.2       & 0.7 \\
2   & Metal       & 0.0     & 0.0       & -0.1 \\
3   & Metal       & 0.0     & 0.0       & 0.0 \\
4   & Rubber       & 9.7     & 5.4       & 5.6 \\
5   & Adhesive       & 2.9     & \textbf{10.2}       & \textbf{41.7} \\
6   & Adhesive       & \textbf{17.6}     & \textbf{9.5}       & \textbf{-100} \\
7   & Adhesive       & 3.7     & 5.1       & 5.1 \\
8   & Film       & -0.4     & -0.7       & -0.9 \\
9   & Film       & \textbf{6.4}     & 3.4       & \textbf{7.5} \\

\botrule
\end{tabular}
\end{center}
\end{table*}

\begin{figure*}[h]
\centering
\includegraphics[width=0.8\textwidth]{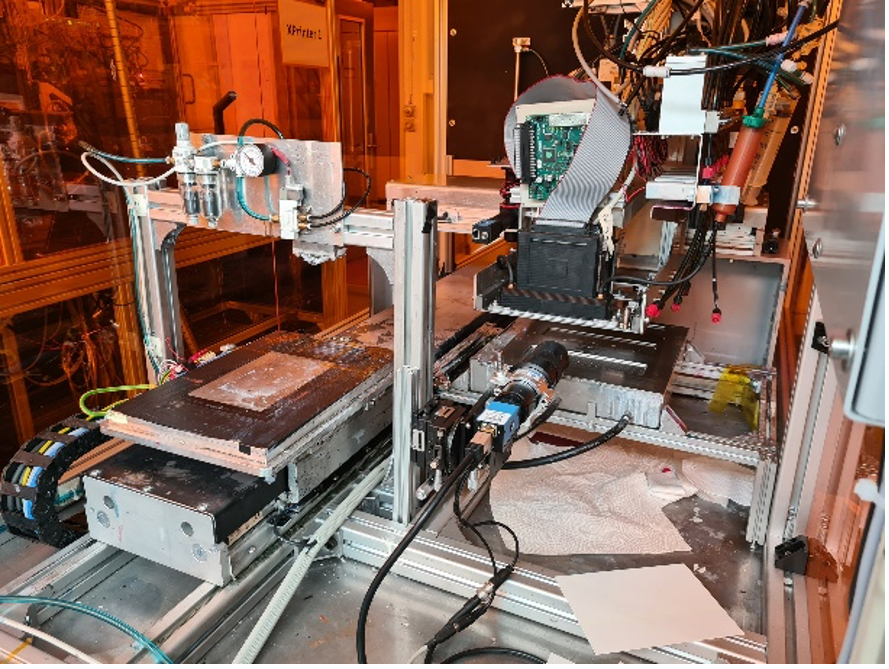}
\caption{Image of the inkjet printing platform employed in this study. The digitally controlled printing stage and ink distribution system are visible on the left, and the dropwatching station is visible on the right.}\label{figS1}
\end{figure*}

\begin{figure*}[h]
\centering
\includegraphics[width=0.8\textwidth]{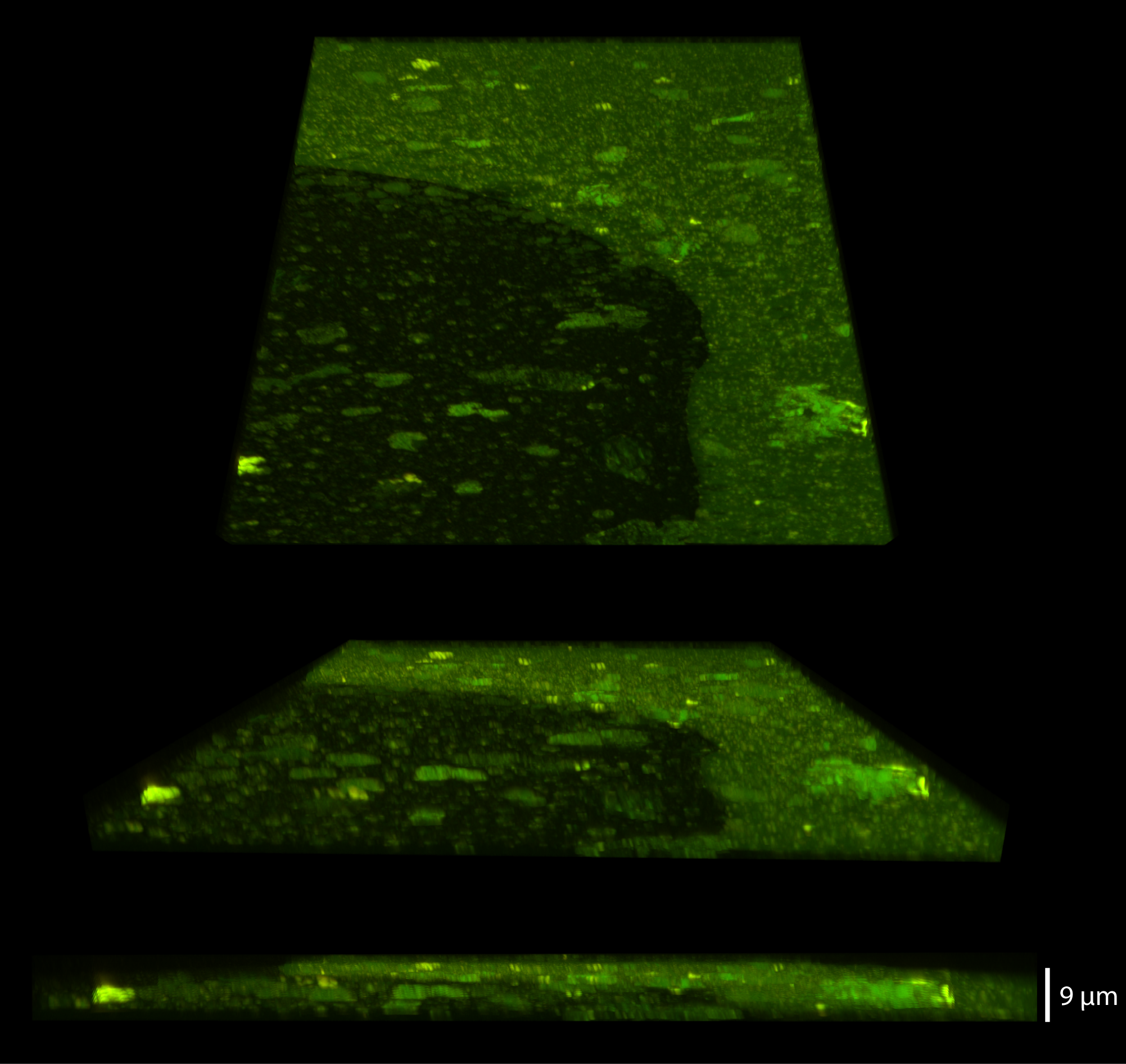}
\caption{Combined-channel 3D CLSM image of an interfacial region between inks A and B, shown from three different viewing angles. These images revealed the dropcast film to be 9 $\mu$m thick.}\label{figS2}
\end{figure*}

\end{appendices}

\end{document}